\documentstyle[12pt,epsfig,axodraw,a4]{article} 
\def\bild#1#2{    
        \vspace*{-5mm}
        \begin{center}
        \begin{math}
        \epsfxsize#2cm
        \epsffile{#1}
        \end{math}
        \end{center}  }
\newcommand{\vs}{\vspace{-0.25cm}}
\begin{document} 

\begin{center}
\large{\bf Spin-orbit coupling in nuclei and realistic nucleon-nucleon 
potentials}

\bigskip

N. Kaiser\\

\medskip

{\small Physik Department T39, Technische Universit\"{a}t M\"{u}nchen,
    D-85747 Garching, Germany}

\end{center}

\medskip

\begin{abstract}
We analyze the spin-orbit coupling term in the nuclear energy density 
functional in terms of a zero-range NN-contact interaction and finite-range
contributions from two-pion exchange. We show that the strength of the
spin-orbit contact interaction as extracted from high-precision nucleon-nucleon
potentials is in perfect agreement with that of phenomenological Skyrme 
forces employed in non-relativistic nuclear structure calculations. Additional
long-range contributions from chiral two-pion exchange turn out to be
relatively small. These explicitly density-dependent contributions reduce
the ratio of the isovector to the isoscalar spin-orbit strength significantly 
below the Skyrme value 1/3. We perform a similar analysis for the 
strength function of the $(\vec \nabla \rho)^2$-term and find values
not far from those of phenomenological Skyrme parameterizations.  

\end{abstract}

\bigskip



\bigskip 
The microscopic understanding the dynamical origin of the strong nuclear 
spin-orbit force is still one of the key problems in nuclear physics. The 
analogy with the spin-orbit interaction in atomic physics gave the hint that it
could be a relativistic effect. This idea has lead to the construction of the 
scalar-vector mean-field models for nuclear structure calculations 
\cite{walecka,ringreview}. In these models the nucleus is described as a 
collection of independent Dirac quasi-particles moving in self-consistently 
generated scalar and vector mean-fields. The footprints of relativity become 
visible through the large nuclear spin-orbit coupling which emerges in that 
framework naturally from the interplay of two strong and counteracting
(scalar and  vector) mean-fields. The corresponding many-body calculations are
usually carried out in the Hartree approximation, ignoring exchange terms as 
well as the negative-energy Dirac-sea. The nucleon-nucleon interaction
introduced into these models is to be considered as an effective one that is
tailored to properties of finite nuclei but not constrained by the observables
of free NN-scattering. For recent ideas about how the large Lorentz scalar and
vector mean-fields can be linked to the condensate structure of the
QCD-vacuum, see ref.\cite{finelli}.  

On the other hand phenomenological Skyrme forces have been (and still are) 
used extensively for the non-relativistic description of nuclei. In
that approach the nuclear spin-orbit coupling is introduced by hand through a 
suitable NN-contact interaction (i.e. a spin-orbit delta-force) with an 
adjustable strength parameter $W_0$. Although there is a wide spread in the 
(other) parameters of the about hundred available Skyrme forces the one 
related to the spin-orbit coupling turns out to be rather stable with a value 
around $W_0 \simeq 120\,$MeVfm$^5$. In fact one may even ignore the concept of
an underlying effective (zero-range) two-body interaction and just take the
emerging (parameterized) energy density functional as the true starting point
for nuclear structure calculations within the self-consistent mean-field 
approximation. Such an interpretation is then in line with the energy 
functional approach to many-body problems of Kohn and Sham. 

The more basic Dirac-Brueckner approach of refs.\cite{brueckner1,brueckner2}
solves a relativistically improved Bethe-Goldstone equation with a realistic
nucleon-nucleon interaction parameterized in terms of one-boson exchange
potentials. This Lorentz-covariant approach is able to describe properties
of nuclear matter and finite nuclei without any new adjustable parameter. In 
particular, the spin-orbit splittings of medium and heavy nuclei are correctly
reproduced (in local density approximation). A drawback of the method is that 
the underlying NN-interaction has been constructed without the explicit use of 
chiral symmetry. In a relativistic Hartree-description there is a close 
connection between the mean-field potential and the spin-orbit one, since they
differ only by the sign of the vector-meson exchange contribution, while in 
non-relativistic approaches there is no connection at all. Unfortunately there
is no direct experimental hint for either possibility. Despite all its known 
successes the fully relativistic treatment of the nuclear many-body problem 
seems unnatural in view of the small ratio $(k_{f0}/M)^2 \simeq 0.08$, with 
$k_{f0}\simeq 263\,$MeV the Fermi momentum of equilibrated nuclear matter and
$M=939\,$MeV the (free) nucleon mass.

The purpose of the present paper is to show that the spin-orbit interaction 
relevant for the structure of (medium and heavy) nuclei is quantitatively 
consistent with that relevant for low-energy elastic NN-scattering. We
demonstrate that the short-range spin-orbit interaction as extracted from 
various high-precision nucleon-nucleon potentials (which accurately fit all 
NN-phase shifts and mixing angles below the NN$\pi$-threshold) is in perfect 
agreement with the one needed in non-relativistic nuclear structure
calculations (using the Skyrme phenomenology). Additional long-range 
spin-orbit couplings generated by chiral two-pion exchange turn out to be
relatively small. These explicitly density dependent effects reduce favorably 
the ratio of the isovector to the isoscalar spin-orbit coupling strength. We 
perform also a similar analysis of the $(\vec \nabla \rho)^2$-term and find 
reasonable values for its strength function (at densities $\rho> 
0.05\,$fm$^{-3}$).

Let us begin with writing down the explicit form of the spin-orbit coupling
term in the nuclear energy density functional:
\begin{equation} {\cal E}_{\rm so}[\rho_p,\rho_n,\vec J_p,\vec J_n] = F_{\rm
so}(k_f)\, \vec\nabla  \rho\cdot\vec J + G_{\rm so}(k_f)\,\vec \nabla  \rho_v 
\cdot\vec J_v \,, \end{equation} 
where the sums $\rho=\rho_p+\rho_n$, $\vec J= \vec J_p+\vec J_n$ and
differences  $\rho_v=\rho_p-\rho_n$, $\vec J_v= \vec J_p-\vec J_n$ of proton
and neutron quantities have been introduced.
\begin{equation} \rho_{p,n}(\vec r\,) = {k_{p,n}^3(\vec r\,) \over 3\pi^2} = 
\sum_{\alpha \in \rm occ} \Psi^{(\alpha)
\dagger}_{p,n}( \vec r\,)\Psi^{(\alpha)}_{p,n}( \vec r\,)\,,\end{equation} 
denote the local proton and neutron densities which we have rewritten in terms
of the corresponding (local) proton and neutron Fermi-momenta $k_{p,n}(\vec
r\,)$ and expressed as sums over the occupied single-particle orbitals
$\Psi^{(\alpha)}_{p,n}( \vec r\,)$. The spin-orbit densities of the protons and
neutrons are defined similarly: 
\begin{equation} \vec J_{p,n}(\vec r\,)=\sum_{\alpha \in \rm occ}\Psi^{(\alpha)
\dagger}_{p,n}(\vec r\,)i\, \vec \sigma \times \vec \nabla\Psi^{(\alpha)}_{p,n
}( \vec r\,) \,. \end{equation} 
Furthermore, $F_{\rm so}(k_f)$ and $G_{\rm so}(k_f)$ in eq.(1) denote the 
density dependent isoscalar and isovector spin-orbit strength functions. In
Skyrme parameterizations \cite{reinhard} these are density-independent 
constants, $F_{\rm so} (k_f)=3G_{\rm so}(k_f)=3W_0/4\simeq 90\,$MeVfm$^5$, 
determined by one single spin-orbit force parameter $W_0$. 

The starting point for the construction of an explicit nuclear energy density 
functional ${\cal E}_{\rm so}[\rho_p,\rho_n,\vec J_p,\vec J_n]$ is the bilocal 
density-matrix as given by a sum over the occupied energy eigenfunctions: 
$\sum_{\alpha\in \rm occ}\Psi^{(\alpha)}_{p,n}( \vec r -\vec a/2)\Psi^{(\alpha
) \dagger}_{p,n}(\vec r +\vec a/2)$. According to Negele and Vautherin 
\cite{negele} it can be expanded in relative and center-of-mass coordinates, 
$\vec a$  and $\vec r$, with expansion coefficients determined by purely local
quantities (nucleon density, kinetic energy density and spin-orbit density). 
As outlined in section 2 of ref.\cite{efun} the Fourier-transform of the (so 
expanded) density-matrix defines in momentum-space a medium-insertion for the 
inhomogeneous many-nucleon system. It is straightforward to generalize this 
construction to the isospin-asymmetric situation of different proton and 
neutron local densities, $\rho_{p,n}(\vec r\,)$ and $\vec J_{p,n}(\vec r\,)$. 
We display here only that part of the medium-insertion $\Gamma(\vec p,\vec q\,
)$ which is actually relevant for the diagrammatic calculation of the isoscalar
and isovector spin-orbit terms defined in eq.(1): 
\begin{eqnarray} \Gamma(\vec p,\vec q\,)& =& \int d^3 r \, e^{-i \vec q \cdot
\vec r}\,\bigg\{ {1+\tau_3 \over 2}\,\theta(k_p-|\vec p\,|) +{1-\tau_3 \over 2}
\,\theta(k_n-|\vec p\,|) \nonumber \\ && +{\pi^2 \over 4k_f^4}
\Big[\delta(k_f-|\vec p\,|) -k_f \,\delta'(k_f-|\vec p\,|) \Big]\, (\vec \sigma
\times \vec p\,) \cdot(\vec J+\tau_3\,\vec J_v) \bigg\}\,.  \end{eqnarray}
The double line in Fig.\,1 symbolizes this medium insertion together with the
assignment of the out- and in-going nucleon momenta $\vec p \pm \vec q/2$. The
momentum transfer $\vec q\,$ is provided by the Fourier components of the
inhomogeneous matter distributions $\rho_{p,n}(\vec r\,)$ and $\vec J_{p,n}(
\vec r\,)$. 

Next, we write down the lowest order four-nucleon contact-coupling which
generates a spin-orbit interaction: 
\begin{eqnarray} {\cal L}_{NN}^{(\rm so)} &=& -i {C_5\over 8} \Big\{(N^\dagger
\vec \nabla N) \cdot (\vec \nabla N^\dagger \times \vec \sigma N) +  (\vec
\nabla N^\dagger N) \cdot (N^\dagger \vec \sigma\times  \vec \nabla N)
\nonumber \\ && \quad\qquad -(N^\dagger N) \,(\vec \nabla N^\dagger \vec
\sigma\times \vec \nabla N)  + (N^\dagger  \vec \sigma N) \cdot (\vec \nabla
N^\dagger \times \vec \nabla N) \Big\} \,.  \end{eqnarray}
This Lagrangian, together with the notation $C_5$ for the coupling constant, 
has been taken over from ref.\cite{evgeni}. The somewhat lengthy expression in
eq.(5) is dictated by Galilei invariance as will become clear immediately. The 
contribution of the contact-Lagrangian ${\cal L}_{NN}^{(\rm so)}$ to the
T-matrix of the scattering process $N(\vec p_1)+ N(\vec p_2)\to N(\vec
p_1\,\!\!')+ N(\vec p_2\,\!\!')$ reads:
\begin{equation} -i{C_5 \over 8}\, (\vec \sigma_1+\vec \sigma_2)\cdot [(\vec
p_1\,\!\!'-\vec  p_2\,\!\!') \times (\vec p_1-\vec p_2)]= -i{C_5 \over 4}\, 
(\vec \sigma_1+\vec \sigma_2)\cdot [(\vec p_1\,\!\!'-\vec  p_1) \times 
(\vec p_1-\vec p_2)]\,, \end{equation} 
where Galilei invariance is now manifest, since only differences of momenta 
occur. $\vec \sigma_{1,2}$ denote the conventional spin-operators of the two
nucleons. Furthermore, by comparing with the analogous T-matrix element of the
Skyrme spin-orbit interaction proportional to $W_0$ (see eq.(4.105) in 
ref.\cite{ringschuck}) one finds the relation $C_5 = 2W_0$ between coupling
parameters. As a side remark we note that the spin-orbit term which arises
from heavy scalar and vector boson exchange between nucleons has the property
of Galilei invariance only if the respective ratios of mass to coupling 
constant are equal: $m_S/g_S=m_V/g_V$. The corresponding contribution to the 
strength parameter $C_5$ reads in that case: $C_5^{(S+V)}=2(1+\kappa_V)(g_V/M 
m_V)^2$ where $M=939\,$MeV stands for (average) the nucleon mass and
$\kappa_V$ denotes the tensor-to-vector coupling ratio of the heavy vector
boson (e.g. the $\omega(782)$-meson). 

In the work of Epelbaum et al. \cite{evgeni} numerical values are given for 
the so-called spectroscopic low-energy constants which characterize the 
short-range part of the nucleon-nucleon potential in certain low partial waves
(S- and P-waves). The spin-orbit strength parameter $C_5=2W_0$ of interest 
here is determined by the following linear combination of the $^3\!P$-wave 
low-energy constants:  
\begin{equation}  C_5 = {1\over 16\pi} \Big[2C(^3\!P_0)+3C(^3\!P_1)-5C(^3\!P_2
)\Big]  \,.  \end{equation}
This simple relation is central to our work since it allows to connect the 
phenomenological Skyrme parameter $W_0=C_5/2$ with the low-energy dynamics of
elastic NN-scattering. 

\medskip

\bild{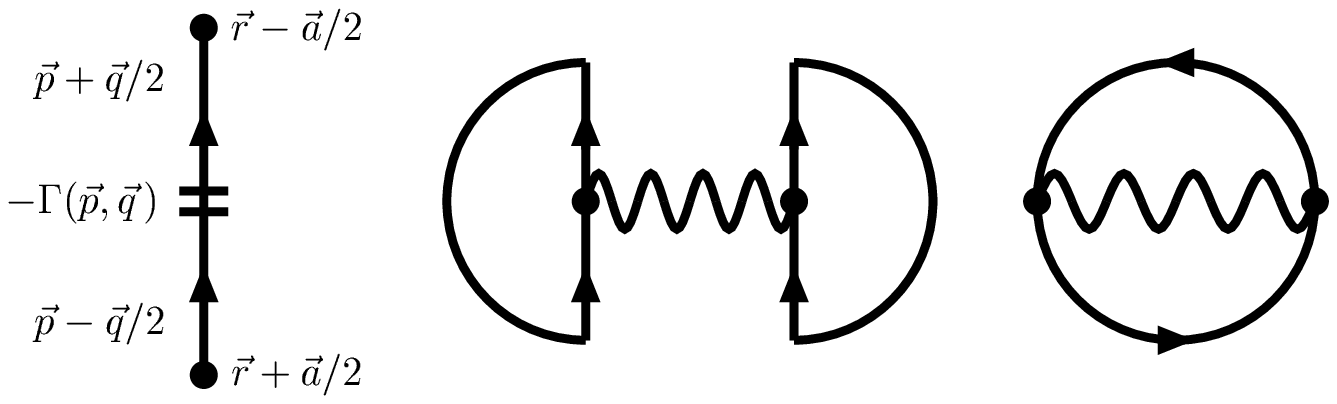}{12}
\vspace{-0.5cm}
{\it Fig.\,1: Left: The double line symbolizes the medium insertion $\Gamma(
\vec p,\vec q\,)$ defined by eq.(4). Next shown are generic two-body Hartree 
and Fock diagrams. Their combinatoric factor is 1/2. The wiggly line
symbolizes the spin-orbit NN-interaction.} 

\medskip

We continue with the general form of the spin-orbit part in the NN T-matrix 
(using the sign-convention of ref.\cite{nnpap}). For the scattering process 
$N(\vec p\,) +N(-\vec p\,)\to N(\vec p+\vec q\,)+N(-\vec p-\vec q\,)$ in the  
center-of-mass frame it reads:
\begin{equation} {\cal T}_{NN}^{(\rm so)} = \Big[ V_{\rm so}(q) + \vec \tau_1 
\cdot \vec \tau_2 \, W_{\rm so}(q)\Big] \, i (\vec \sigma_1+\vec \sigma_2)
\cdot (\vec q \times \vec p\,) \,, \end{equation}
Here, $\vec q$ is the momentum transfer between both nucleons and $V_{\rm so}
(q)$ and $W_{\rm so}(q)$ denote the isoscalar and isovector spin-orbit 
NN-amplitudes, respectively. With the help of the medium insertion 
$\Gamma(\vec p, \vec q\,)$ in eq.(4) one can now calculate diagrammatically
the nuclear energy density functional ${\cal E}_{\rm so}[
\rho_p,\rho_n,\vec J_p,\vec J_n]$. From the Hartree and Fock diagrams in 
Fig.\,1 one obtains the following formulas for two-body contributions to the
isoscalar and isovector spin-orbit strength functions:       
\begin{equation} F_{\rm so}(k_f)^{(\rm 2-body)} = -{1\over 6} \bigg\{ 3
V_{\rm so}(0)+ V_{\rm so}(2k_f)+3 W_{\rm so}(2k_f)+ \int_0^1 dx\,x \Big[ 
V_{\rm so}(2xk_f) + 3W_{\rm so}(2x k_f) \Big] \bigg\} \,, \end{equation} 
\begin{equation} G_{\rm so}(k_f)^{(\rm 2-body)} = {1\over 6} \bigg\{  
W_{\rm so}(2k_f)-V_{\rm so}(2k_f)-3W_{\rm so}(0) + \int_0^1 dx\,x \Big[ 
W_{\rm so}(2xk_f) -V_{\rm so}(2x k_f) \Big] \bigg\} \,. \end{equation} 
Here, the terms $-V_{\rm so}(0)/2$ and $-W_{\rm so}(0)/2$ belong to the 
Hartree diagram (with two closed nucleon lines) while the remaining ones
summarize the contribution from the Fock diagram (having just one closed 
nucleon line). After the obvious identification $V_{\rm so}(q)^{(\rm ct)} = 
-C_5/2$ (compare eq.(6) with eq.(8)) one derives the following contribution 
of the four-nucleon contact-vertex to the spin-orbit strength functions:
\begin{equation} F_{\rm so}(k_f)^{(\rm ct)} = {3C_5\over 8} \,, \qquad
G_{\rm so}(k_f)^{(\rm ct)} = {C_5\over 8} \,. \end{equation} 
The fixed ratio of isovector-to-isoscalar spin-orbit strength $G_{\rm
so}(k_f)/F_{\rm so}(k_f)= 1/3$ is a consequence of the Pauli exclusion
principle for a zero-range spin-orbit interaction. 

The long-range contributions to the spin-orbit NN-potential arise naturally 
from two-pion exchange between nucleons in the form of a relativistic
$1/M$-correction. The corresponding lowest order $2\pi$-exchange triangle and 
box diagrams have been evaluated in section 4.2 of ref.\cite{nnpap}. Inserting 
the corresponding analytical expressions for $V_{\rm so}(q)$ and $W_{\rm so}
(q)$ (see eqs.(22,23) in ref.\cite{nnpap}) into the master formulas eqs.(9,10)
one gets the following contributions from chiral $2\pi$-exchange to the 
spin-orbit strength functions:  
\begin{equation} F_{\rm so}(k_f)^{(2\pi)} = {g_A^2 m_\pi \over \pi M(4f_\pi)^4}
\bigg\{ {10\over 3} - {3g_A^2 \over 2} +{4-3g_A^2 \over 6u^2} \ln(1+u^2)
+ \bigg[ {2\over u}(g_A^2-2) -{8u \over 3} \bigg] \arctan u
\bigg\} \,, \end{equation}  
\begin{eqnarray} G_{\rm so}(k_f)^{(2\pi)} &=& {g_A^2 m_\pi \over 9\pi M(4f_\pi
)^4} \bigg\{ {53g_A^2 \over 2}- 10+{7g_A^2-4 \over 2u^2} \ln(1+u^2) \nonumber
\\ && + \bigg[  {6\over u}(2-5g_A^2) +8u(1-4g_A^2)\bigg] \arctan u \bigg\} \,,
\end{eqnarray} 
with the abbreviation $u= k_f/m_\pi$ where $m_\pi = 135\,$MeV stands for the
(neutral) pion mass. As usual $f_\pi = 92.4\,$MeV denotes the weak pion decay 
constant and we choose the value $g_A = 1.3$ for the nucleon axial vector 
coupling constant. Note that we have normalized the density dependent 
expressions in eqs.(12,13) to the value zero at zero density $(k_f= 0)$. This 
way one eliminates automatically all (regularization dependent) short-range 
contributions which conceptually belong to the low-energy constant $C_5$. As 
stressed in ref.\cite{evgeni} such a separation of long and short-distance
dynamics is necessary in order to be able to compare with results extracted
from realistic NN-potentials. At the next order in the small momentum
expansion there are $2\pi$-exchange diagrams with one chiral $\pi\pi
NN$-contact vertex. The corresponding spin-orbit NN-amplitudes have been 
written down in eqs.(13,15,16) of ref.\cite{2loop}. After performing the
necessary integration and subtraction at $k_f=0$ one gets the following
$2\pi$-exchange contributions to the spin-orbit strength functions:
\begin{eqnarray} F_{\rm so}(k_f)^{(2\pi-c_j)} &=& {m_\pi^2 \over 3\pi^2 M(4
f_\pi)^4} \bigg\{ 20u^2\Big[ c_4+g_A^2(5c_4-2c_2) \Big] \ln{m_\pi\over\Lambda} 
\nonumber \\ && -29 c_4 +g_A^2(58c_2-61c_4) +u^2 \Big[g_A^2(2c_2-5c_4)- c_4 
\Big] \nonumber \\ &&+ \bigg[ {2\over u} \Big(13c_4 +g_A^2(29c_4-26c_2) \Big) +
20u \Big( c_4 +g_A^2(5c_4-2c_2) \Big) \bigg]\sqrt{1+u^2} \nonumber \\ &&
\times \ln(u+ \sqrt{1+u^2} )+ {3\over u^2} \Big[c_4 +g_A^2(c_4-2c_2)
\Big] \ln^2(u+\sqrt{1+u^2} ) \bigg\} \,, \end{eqnarray} 
\begin{eqnarray} G_{\rm so}(k_f)^{(2\pi-c_j)} &=& {m_\pi^2 \over 9\pi^2 M(4
f_\pi)^4} \bigg\{-20u^2\Big[ c_4+g_A^2(5c_4+6c_2) \Big] \ln{m_\pi\over\Lambda} 
\nonumber \\ && +29 c_4 +g_A^2(174c_2+61c_4) +u^2 \Big[c_4+g_A^2(6c_2+5c_4) 
\Big] \nonumber \\ &&- \bigg[ {2\over u} \Big(13c_4 +g_A^2(78c_2+29c_4) \Big) +
20u \Big( c_4 +g_A^2(5c_4+6c_2) \Big) \bigg]\sqrt{1+u^2} \nonumber \\ &&
\times \ln(u+ \sqrt{1+u^2} )- {3\over u^2} \Big[c_4 +g_A^2(c_4+6c_2)
\Big] \ln^2(u+\sqrt{1+u^2} ) \bigg\} \,. \end{eqnarray} 
The two occurring low-energy constants $c_2= 3.2\,$GeV$^{-1}$ and $c_4=
3.4\,$GeV$^{-1}$ \cite{buett} represent mainly effects from (single) virtual 
$\Delta(1232)$-isobar excitation. In these expressions we have also introduced
$k_f^2$-terms proportional to the chiral logarithm $\ln(m_\pi/\Lambda)$ in 
order to guarantee finite results for $F_{\rm so}(k_f)^{(2\pi-c_j)}$ and 
$G_{\rm so}(k_f)^{(2\pi-c_j)}$ in the strict chiral limit $m_\pi =0$.   
  
\begin{table}[hbt]
\begin{center}
\begin{tabular}{|c|cccccc|}
    \hline
    Skyrme force & SIII & SkM & SkP & Sly4-7 & MSk1-6 & SkI1-5 \\ 
\hline $3W_0/4$ [MeVfm$^5$]  & 90.0 & 97.5 & 75.0 & 93.2 & 87.6 & 92.7 \\
    \hline 
\end{tabular}
\end{center}
{\it Table I: Numerical values of the spin-orbit strength parameter 
$3W_0/4$ for various phenomenological Skyrme forces.}
\vskip-1.0cm
\end{table}
\begin{table}[hbt]
\begin{center}
\begin{tabular}{|c|ccccccc|}
    \hline
   NN-potential & Bonn-B & CD-Bonn$^*$ & Nijm-93 & Nijm-I & Nijm-II$^*$ &
    AV-18$^*$ & NNLO \\ \hline $3C_5/8$ [MeVfm$^5$]  & 80.3 & 89.6 & 79.9 &
    82.4 & 87.7 & 88.9 & 73 $\cdots$ 92 \\     \hline
\end{tabular}
\end{center}
{\it Table II: Numerical values of the short-range spin-orbit strength
parameter $3C_5/8$ for various nucleon-nucleon potentials. The so-called
high-precision potentials are marked by an asterisk.} 
\end{table}

Now we can turn to numerical results. In Table I we have collected values for
the isoscalar spin-orbit strength $3W_0/4$ of various phenomenological Skyrme
forces, SIII \cite{s3}, SkM \cite{skm}, SkP \cite{skp}, Sly \cite{sly},
MSk \cite{pearson} and SkI \cite{b4paper}. In the cases of Sly, MSk and SkI we
have performed averages over several (slightly different) parameter sets 
Sly4-7, MSk1-6 and SkI1-5. One sees that the entries in Table I scatter
somewhat around a mean value of about $3W_0/4 \simeq 90\,$MeVfm$^5$ which can
therefore be considered as the empirical spin-orbit strength relevant for
nuclear structure. In Table II we give the equivalent spin-orbit strength
parameter $3C_5/8$ as extracted from various realistic nucleon-nucleon
potentials. These numbers have been produced with the help of Table IV in
ref.\cite{evgeni} by forming the appropriate linear combination eq.(7) of 
$C(^3\!P_{0,1,2})$. The last entry ''NNLO'' in Table II corresponds to the 
chiral NN-potential of Epelbaum et al. \cite{evgeni} where the cut-off 
$\Lambda$ has been varied  between 0.5 and 0.6\,GeV (see Table I in 
ref.\cite{evgeni}). It is astonishing to observe that these realistic 
NN-potentials which are essentially adjusted to empirical low-energy 
NN-scattering data give numbers close to those of the phenomenological Skyrme 
forces. This holds in particular for the so-called high-precision 
NN-potentials CD-Bonn \cite{cdbonn}, Nijm-II \cite{nijm} and 
AV-18 \cite{av18} (marked by an asterisk in Table II) where $3C_5/8$ comes out 
very close to $3C_5/8\simeq 90\,$MeVfm$^5$. Therefore one can conclude that 
the strength of the spin-orbit interaction necessary for nuclear structure is 
perfectly consistent with that needed to describe low-energy NN-scattering 
data. This observation is the one of the main results of the present work.

The recently constructed universal low-momentum nucleon-nucleon potential 
$V_{\rm low-k}$ offers another possibility to extract the short-range
spin-orbit parameter $3C_5/8$. From the curvatures of the (diagonal) 
$^3\!P$-wave potentials shown in Figs.\,4,18 of ref.\cite{vlowk} we get after 
multiplying the relevant linear combination $2(^3\!P_0) + 3(^3\!P_1)
-5(^3\!P_2)$ with the conversion factor $3\pi/8M$ the value $3C_5/8 = 79\,
$MeVfm$^5$ from $V_{\rm low-k}$. This somewhat smaller value (in comparison to
the high-precision potentials) finds its explanation in the fact that the
''full'' NN-potential and $V_{\rm low-k}$ differ by some local contact-terms 
\cite{counter}. Indeed  the numerical values of the counter-terms listed in 
Table\,1 of ref.\cite{counter} give rise to an additional a contribution to
$3C_5/8$ of $10.4\, $MeVfm$^5$ which then closes the gap the empirical value 
of $3C_5/8 \simeq 90\,$MeVfm$^5$.

\bigskip

\bild{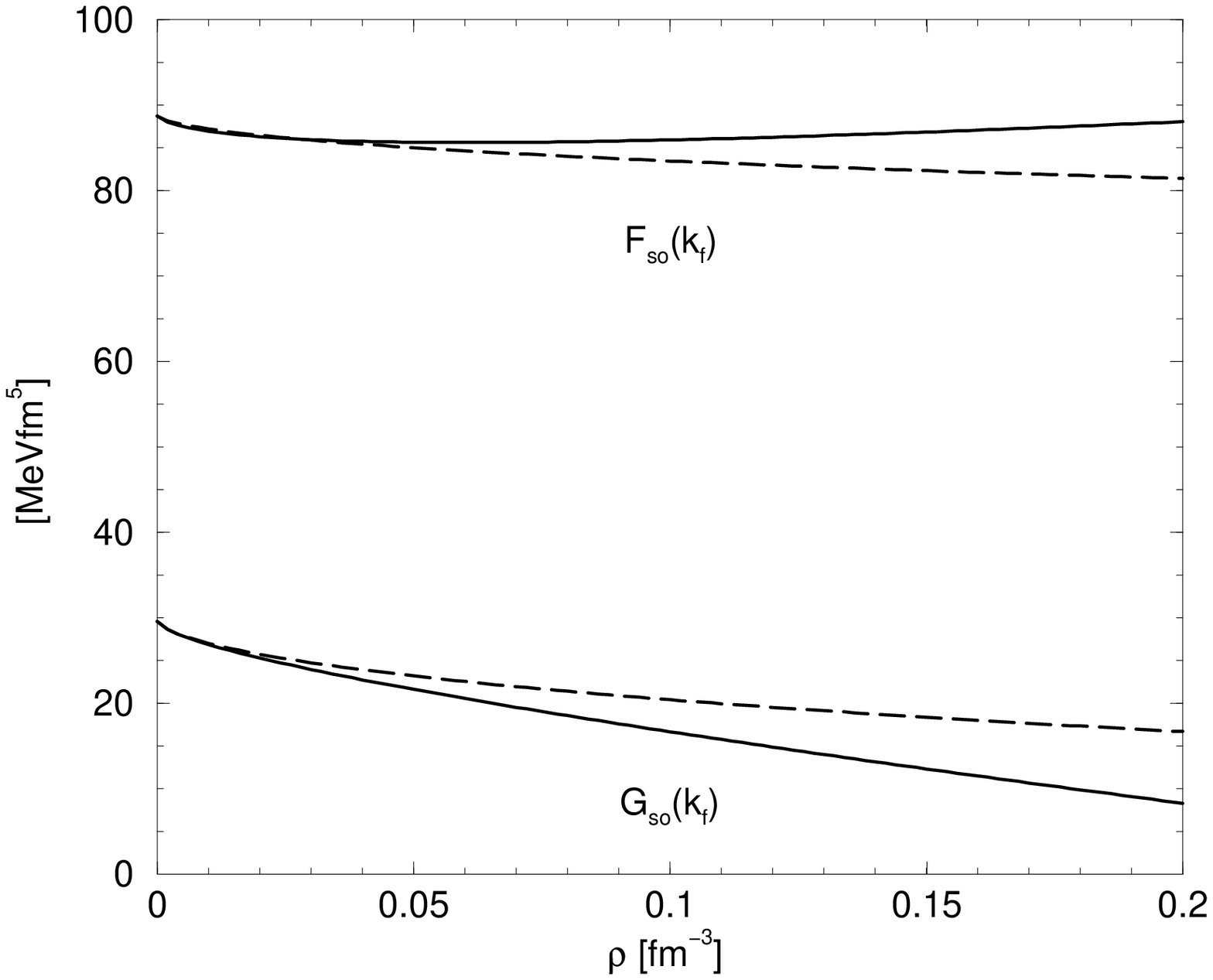}{12}
\vskip-0.5cm
{\it Fig.\,2: The isoscalar and isovector spin-orbit strength functions 
$F_{\rm so}(k_f)$ and $G_{\rm so}(k_f)$ versus the nucleon density $\rho=
2k_f^3/3\pi^2$. The dashed curves show the result of the spin-orbit 
contact-coupling $C_5$ and irreducible two-pion exchange eqs.(12,13). The 
full curves include in addition the two-pion exchange contributions
eqs.(14,15) proportional to $c_{2,4}$. The scale in the chiral logarithm is 
chosen as $\Lambda = 0.5\,$GeV.}

\bigskip

In Fig.\,2 we show the isoscalar and isovector spin-orbit strength functions
$F_{\rm so}(k_f)$ and $G_{\rm so}(k_f)$ as a function of the nucleon density
$\rho= 2k_f^3/3\pi^2$. The dashed lines result from the average value $3C_5/8 
= 88.7\,$MeVfm$^5$ of the three high-precision NN-potentials  and the 
lowest order irreducible $2\pi$-exchange contributions eqs.(12,13). The full 
lines include in addition the next-to-leading order $2\pi$-exchange
contributions proportional to the low-energy constants $c_2 = 3.2\,$GeV$^{-1}$
and $c_4 = 3.4\,$GeV$^{-1}$ \cite{buett}. The scale $\Lambda$ in the chiral
logarithm $\ln(m_\pi /\Lambda)$ has been chosen equal to the momentum cut-off 
$\Lambda = 0.5\,$GeV of ref.\cite{evgeni}. In the case of the isoscalar
spin-orbit strength function $F_{\rm so}(k_f)$ one finds that the two-pion
exchange effects are relatively small and that they also cancel each other to 
a large extent. In contrast to this behavior does the isovector spin-orbit 
strength function $G_{\rm so}(k_f)$ get reduced with increasing density. At 
nuclear matter saturation density $\rho_0 \simeq 0.16\,$fm$^{-3}$ the ratio  
$G_{\rm so}(k_{f0})/F_{\rm so}(k_{f0})$ is about $1/8$, substantially less
than the Skyrme value $1/3$. As pointed out in ref.\cite{b4paper} such a
reduction of the isovector spin-orbit mean-field  (proportional to $\vec
\nabla \rho_n - \vec \nabla \rho_p$) is favorable for an accurate description 
of isotope shifts in the Pb region. It seems that the isospin dependence of 
the spin-orbit interaction is a property for which two-pion exchange induced
effects can play a role in nuclear structure.   

For the sake of completeness we also note that even the one-pion exchange Fock
diagram gives rise to non-vanishing spin-orbit strength functions:
\begin{equation} F_{\rm so}(k_f)^{(1\pi)}= -3 G_{\rm so}(k_f)^{(1\pi)} =
{g_A^2 \over (16 M f_\pi)^2} \Big[ u^{-2} \ln(1+4u^2) -4 \Big]
\,.\end{equation} 
In order to arrive at this result one has to expand the relativistic 
pseudovector $\pi NN$-vertex up to order $1/M^2$. As expected the 
$1\pi$-exchange contributions to the spin-orbit strengths are negligibly
small. For example, at a density of $\rho_0/2 \simeq 0.08\,$fm$^{-3}$ one gets 
$F_{\rm so}^{(1\pi)}(k_f) \simeq -0.79\,$MeVfm$^5$ which corresponds in 
magnitude to less than $1\%$ of the empirical value $90\,$MeVfm$^5$.

Finally, we perform the same analysis for the $(\vec \nabla \rho)^2$-term in
the nuclear energy density functional: $F_\nabla(k_f) \,(\vec \nabla
\rho)^2$. Its strength function $F_\nabla(k_f)$ is also decomposed into
contributions from zero-range NN-contact interactions and long-range one- and 
two-pion exchange. From the complete set of four-nucleon contact-couplings
written down in ref.\cite{evgeni} one gets:      
\begin{eqnarray}  F_{\nabla}(k_f)^{(\rm ct)}&=& {1\over 32}
(14C_1+C_2-6C_3-3C_4-2C_6-C_7) \nonumber \\ &=&  {3\over 512 \pi}
\Big[6C(^1\!S_0) +6C(^3\!S_1)-C(^1\!P_1)- C(^3\!P_0)-3C(^3\!P_1)-5C(^3\!P_2)
\Big]\,,  \end{eqnarray}
where we have reexpressed the relevant linear combination of $C_1,\dots,C_7$
through the so-called spectroscopic low-energy constants. In that
representation we obtain from the entries in Table IV of ref.\cite{evgeni} for
the three high-precision NN-potentials CD-Bonn, Nijm-II and AV-18 the values 
$F_{\nabla}^{(\rm ct)}(k_f)= 82.4 \,$MeVfm$^5$, $78.4\,$MeVfm$^5$ and 
$82.5\,$MeVfm$^5$, respectively. The contribution of the one-pion exchange
Fock-diagram to the strength function $F_{\nabla}(k_f)$ reads on the other
hand \cite{efun}: 
\begin{eqnarray}  F_{\nabla}(k_f)^{(1\pi)}&=& {35g_A^2 u^{-7}\over (16 m_\pi
f_\pi)^2} \bigg\{ {7\over 8u} -{21u \over 2}-{u^3\over 6} +{72u^4-90u^2-7
\over 32 u^3} \ln(1+4u^2) \nonumber \\ && +10 \arctan 2u + {m_\pi^2 \over 16
M^2} \bigg[ {3\over 4u} +{193u \over 2}-12u^3 +{4u^5 \over 3} -{96 u^7 \over
35} \nonumber \\ && +20(2u^2-3)\arctan 2u +\bigg( {11 \over 2u} -45u
-{3\over 16 u^3} \bigg) \ln(1+4u^2)\bigg] \bigg\} \,, \end{eqnarray}
where we have included also the relativistic $1/M^2$-correction. The
contributions of (lowest order) irreducible $2\pi$-exchange to the NN-potential
have been evaluated in ref.\cite{nnpap} (see eqs.(14,15) therein). Application
of the density-matrix expansion of Negele and Vautherin \cite{negele} as 
outlined in section 2 of ref.\cite{efun} leads then to the following 
contribution of irreducible $2\pi$-exchange to the strength function  
$F_{\nabla}(k_f)$:   
\begin{eqnarray}  F_{\nabla}(k_f)^{(2\pi)}&=& {u^{-10}\over \pi^2(16 f_\pi)^4}
\bigg\{ \Big[ 245(71g_A^4+10g_A^2-1)+1575u^2 (43g_A^4+6g_A^2-1)\nonumber \\ && 
-840u^4(23g_A^4+2g_A^2-1) \Big] \ln^2(u+\sqrt{1+u^2}) +{2u \over 3} 
 \ln(u+\sqrt{1+u^2})\nonumber \\ && \times \sqrt{1+u^2}\Big[735(1-10g_A^2-71
g_A^4) +5u^2 (31859g_A^4+3502g_A^2-1201) \nonumber \\ && +2u^4(869g_A^4-230
g_A^2-79)+24u^6(2u^2-1)(11g_A^4-10g_A^2-1) \Big]\nonumber \\ &&
+245u^2(71g_A^4 +10g_A^2-1) -{20u^4 \over 3} (23479g_A^4+2801g_A^2-800)
\nonumber \\ && +{u^6 \over 3}(551 + 470g_A^2-7181g_A^4)+4u^8 (27g_A^4-22g_A^2
-5) \nonumber \\ && +{4u^{10}\over  105}(869+7010g_A^2 -6199g_A^4) \bigg\} \,. 
\end{eqnarray} 
Again, in order to have a clean separation between (regularization dependent)
short-distance effects and genuine long-distance contributions we have 
normalized this  expression to $F_{\nabla}(0)^{(2\pi)}=0$. 

\bigskip

\bild{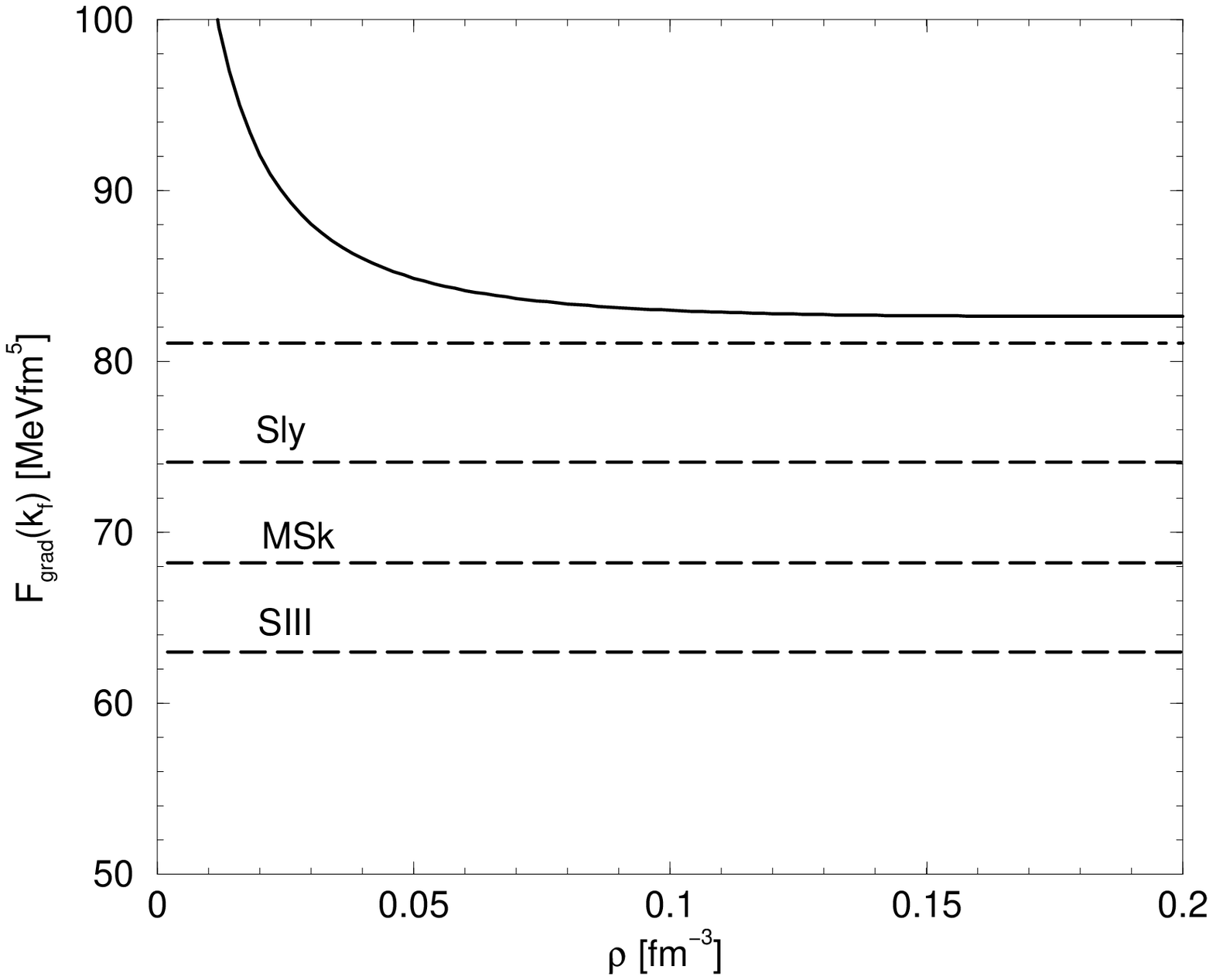}{12}
\vskip-0.5cm
{\it Fig.\,3: The strength function $F_\nabla(k_f)$ related to the $(\vec
\nabla \rho)^2$-term in the nuclear energy density functional versus the 
nucleon density $\rho= 2k_f^3/3\pi^2$. The three horizontal dashed lines show
the constant values  $F_\nabla(k_f)=[9t_1-(5+4x_2)t_2]/64$ of the Skyrme 
forces Sly \cite{sly}, MSk \cite{pearson} and SIII \cite{s3}. The 
dashed-dotted line gives the mean value $F_\nabla(k_f)^{(\rm ct)} =81.1\,
$MeVfm$^5$ of the three high-precision NN-potentials CD-Bonn, Nijm-II and 
AV-18. The full line includes in addition the $1\pi$- and $2\pi$-exchange
contributions eqs.(18,19).} 

\bigskip 

In Fig.\,3 we show the strength function $F_\nabla(k_f)$ versus the nucleon 
density $\rho= 2k_f^3/3\pi^2$. The three horizontal dashed lines represent the 
constant values  $F_\nabla(k_f)=[9t_1-(5+4x_2)t_2]/64$ of the Skyrme 
forces Sly \cite{sly}, MSk \cite{pearson} and SIII \cite{s3}. The 
dashed-dotted line gives the mean value $F_\nabla(k_f)^{(\rm ct)} =81.1\,
$MeVfm$^5$ of the three high precision potentials CD-Bonn, Nijm-II and AV-18. 
The full line includes in addition the $1\pi$- and $2\pi$-exchange
contributions eqs.(18,19). One observes a rough agreement between these 
different curves. For densities $\rho> 0.05\,$fm$^{-3}$ our prediction for 
$F_\nabla(k_f)$ exceeds the mean value of the phenomenological  Sly4-7 forces 
\cite{sly} only by about $15\%$. Below that density a strong rise of the 
strength function $F_\nabla(k_f)$ sets however in. This behavior originates 
from the static $1\pi$-exchange contribution with its inherent chiral 
singularity $m_\pi^{-2}$ which becomes dominant at extremely low densities 
$k_f << m_\pi/2$ \cite{efun}. Our derivation of the strength function
$F_\nabla(k_f)$ is based on the density-matrix expansion of Negele and
Vautherin \cite{negele} which has been found to become inaccurate for low and
nonuniform densities \cite{dobac}. Therefore one should not trust the full 
line in Fig.\,3 below $\rho \leq \rho_0/4 \simeq 0.04\,$fm$^{-3}$. It is also 
evident from Fig.\,3 that the nuclear structure phenomenology does presently 
not constrain the $(\vec \nabla \rho)^2$-term as strongly than as the 
spin-orbit coupling term. The existing Skyrme forces show an appreciable 
variation of the associated strength parameter.       

In summary, we analyzed in this work the spin-orbit coupling term in the 
nuclear energy density functional in terms of a zero-range NN-contact
interaction and finite-range contributions from chiral two-pion exchange. We 
have shown that the strength of the spin-orbit contact-interaction as
extracted from high-precision NN-potentials agrees perfectly with that of
phenomenological Skyrme forces employed in non-relativistic nuclear structure
calculations. The numerical results collected in Table IV of ref.\cite{evgeni}
have been essential in order to establish this fact. Additional long-range 
effects from chiral two-pion exchange seem to play a role for the isospin 
dependence of the spin-orbit coupling. The ratio of the isovector to the 
isoscalar spin-orbit strength gets significantly reduced below the Skyrme 
value 1/3. Such a reduction is favorable for an accurate description of the 
isotope shifts in Pb nuclei \cite{b4paper}. Furthermore, we
have performed the same  analysis for the strength function of the $(\vec
\nabla \rho)^2$-term and found values not far from those of phenomenological 
Skyrme parameterizations.

\end{document}